\documentstyle[12pt]{article}

\begin{document}
\begin{titlepage}

\title{Accelerated observers and gravitational radiation}

\author{J. W. Maluf$\,^{*}$ \\
Instituto de F\'{\i}sica, \\
Universidade de Bras\'{\i}lia\\
C. P. 04385 \\
70.919-970 Bras\'{\i}lia DF, Brazil\\}
\date{}
\maketitle

\begin{abstract}

We evaluate the energy-momentum of the gravitational field of a
Schwarzschild black hole of mass $M$ in the frame of a moving
observer that asymptotically undergoes a Lorentz boost. The analysis
is carried out in the framework of the teleparallel equivalent of
general relativity (TEGR). We find that the total expression for the
energy-momentum of the gravitational field is similar to the usual
relativistic expression for the energy-momentum four-vector of a
particle of inertial mass $M$ under a Lorentz boost in flat
space-time. Moreover we conclude that if the observer accelerates
with respect to the black hole he will experience gravitational
energy radiation, in similarity to the expected radiation of an
accelerated charged particle in electrodynamics. We show that the
increase of the mass of the black hole by the usual factor $\gamma$
as observed in the moving frame, which is a typical feature of
special relativity if the black hole is considered asymptotically
as a body of mass $M$, is due to gravitational radiation.

\end{abstract}
\thispagestyle{empty}
\vfill
\par
\bigskip
\noindent (*) e-mail: wadih@fis.unb.br\par
\end{titlepage}

\bigskip
\bigskip

The two standard notions of gravitational energy are the total
energy of the space-time, known as the ADM (Arnowitt-Deser-Misner)
energy \cite{ADM}, evaluated on a spacelike hypersurface of a
space-time that is asymptotically flat, and the Bondi energy
\cite{Bondi} that describes the mass of radiating systems in
asymptotically flat space-times. The latter is usually defined at
null infinity. Earlier attempts at defining the gravitational energy
were based on pseudotensors \cite{Landau,Weinberg}, which reproduce
the ADM energy by requiring the appropriate boundary
conditions \cite{Goldberg}. In none of these approaches the
characterization of the observer is
explicitly taken into account in the definition of the gravitational
energy. In the case of asymptotically flat space-times it might be
assumed that the observer is static at spacelike infinity, or that
the asymptotically flat coordinate system is adapted to a
hypothetical observer.

The definition of gravitational energy has been addressed
in the realm of the teleparallel equivalent of general
relativity (TEGR) \cite{Maluf1,Maluf2}. Conceptually, 
this approach is reminiscent
of a previous attempt by M\o ller \cite{Mol}, who noticed
that an expression for the gravitational energy density could
be constructed out of the torsion tensor. The torsion tensor cannot
be made to vanish at a point by a coordinate transformation. This
fact refutes the usual argument against the nonlocalizability of the
gravitational energy, which is ordinarily attributed to the
principle of equivalence, and which rests on the reduction of the
metric tensor to the Minkowski metric tensor at a point in space-time
by means of a coordinate transformation. In our opinion the principle
of equivalence plays a totally different role in this respect. The
principle is based on the equality of inertial and gravitational
masses. Alternatively, an accelerated frame can be locally considered
as a rest frame with the addition of a certain gravitational field.
Thus our perception of the strength of the gravitational field
on nearby bodies clearly depends on our reference frame, and so does
the gravitational energy as measured on the same frame. Therefore
the principle of equivalence implies a dependence of the
gravitational energy density on the frame, not on the coordinate
system. Reference frames are described in terms of tetrad fields
\cite{Synge,Aldrovandi,Mashhoon1}, which are the basic field
quantities of the TEGR.

An essential feature of the gravitational energy-momentum
$P^a=(E/c,{\bf P})$
established in the framework of the TEGR \cite{Maluf1} is the
covariance under  global SO(3,1) transformations, in addition
to the invariance
under coordinate transformations of the three-dimensional
spacelike hypersurface. Each configuration of tetrad fields
establishes a reference frame adapted to an observer.
$P^a$ is constructed out of the torsion tensor
$T^a\,_{\mu \nu}=\partial_\mu e^a\,_\nu-\partial_\nu e^a\,_\mu$,
where $e^a\,_\mu$ represents the tetrad field.
(Notation: space-time indices $\mu, \nu, ...$ and SO(3,1)
indices $a, b, ...$ run from 0 to 3. Time and space indices are
indicated according to $\mu=0,i,\;\;a=(0),(i)$).
The flat, Minkowski space-time  metric is fixed by
$\eta_{ab}=e_{a\mu} e_{b\nu}g^{\mu\nu}= (-+++)\;$. The torsion tensor
above is related to the antisymmetric part of the Weitzenb\"ock
connection $\Gamma^\lambda_{\mu\nu}=
e^{a\lambda}\partial_\mu e_{a\nu}$. The curvature tensor constructed
out of this connection vanishes identically). The definition of
$P^a$ arises in the analysis of the Hamiltonian formulation of the
TEGR \cite{Maluf3}. The energy definition $P^{(0)}$ has been applied
to several configurations of the gravitational field
(see Ref. \cite{Maluf1} and references therein).

In this article we will 
establish the tetrad field that describes a moving observer in the
space-time of a Schwarzschild black hole. The frame motion is
characterized asymptotically by a Lorentz boost. We conclude that
$P^a=Mcu^a$, where $M$ is the mass of the black hole, $c$ is the
velocity of light and  $u^a$ is the four-velocity of the frame.
Therefore the resulting expression is precisely the same as the
energy-momentum of a relativistic particle of inertial mass $M$ in
flat space-time, that undergoes a Lorentz boost. Moreover if the
frame velocity is time dependent, the expression for the total
gravitational energy flux measured in the accelerated frame is
nonvanishing, a result that confirms that accelerated bodies in
general relativity radiate \cite{Bicak,Tomimatsu} exactly like
accelerated charged particles in electrodynamics. The present
analysis indicates that the gravitational energy-momentum $P^a$
may provide a unified picture of the concept of mass-energy in the
special and general theory of relativity.

The Lagrangian density for the gravitational field in the TEGR, in
the presence of matter fields, is given by

\begin{eqnarray}
L(e_{a\mu})&=& -k\,e\,({1\over 4}T^{abc}T_{abc}+
{1\over 2} T^{abc}T_{bac} -T^aT_a) -{1\over c}L_m\nonumber \\
&\equiv&-k\,e \Sigma^{abc}T_{abc}-{1\over c}L_m\;,
\label{1}
\end{eqnarray}
where $k=c^3/(16 \pi G)$, $e=\det(e^a\,_\mu)$ and $x^0=ct$. The
tensor $\Sigma^{abc}$ is defined by

\begin{equation}
\Sigma^{abc}={1\over 4} (T^{abc}+T^{bac}-T^{cab})
+{1\over 2}( \eta^{ac}T^b-\eta^{ab}T^c)\;,
\label{2}
\end{equation}
and $T^a=T^b\,_b\,^a$. The quadratic combination
$\Sigma^{abc}T_{abc}$ is proportional to the scalar curvature
$R(e)$, except for a total divergence. $L_m$
represents the Lagrangian density for matter fields. The field
equations for the tetrad field read

\begin{equation}
e_{a\lambda}e_{b\mu}\partial_\nu(e\Sigma^{b\lambda \nu})-
e(\Sigma^{b \nu}\,_aT_{b\nu \mu}-
{1\over 4}e_{a\mu}T_{bcd}\Sigma^{bcd})
\;=\;{1\over {4kc}}eT_{a\mu} \;,
\label{3}
\end{equation}
where $\delta L_m/\delta e^{a\mu}\equiv eT_{a\mu}$
It is possible to prove by explicit calculations that the left hand
side of Eq. (3) is exactly given by ${1\over 2}\,e\,
\lbrack R_{a\mu}(e)-{1\over 2}e_{a\mu}R(e)\rbrack$.

The definition of the gravitational energy-momentum contained within
an arbitrary volume $V$ of the three-dimensional spacelike
hypersurface arises in the Hamiltonian formulation of the
TEGR \cite{Maluf3}. It reads

\begin{equation}
P^a=-\int_V d^3x\, \partial_j \Pi^{aj}\;,
\label{4}
\end{equation}
where $\Pi^{aj}=-4ke\Sigma^{a0j}$ is the momentum canonically
conjugated to $e_{aj}$. As we will discuss later, the expression
above bears no relationship to the ADM energy-momentum. 
Simple algebraic manipulations of Eq. (3)
yield a continuity equation for the gravitational energy-momentum
$P^a$ \cite{Maluf4,Maluf5},

\begin{equation}
{{d P^a} \over {dx^0}} =
-\Phi^a_g -\Phi^a_m\;,
\label{5}
\end{equation}
where

\begin{equation}
\Phi^a_g=k\int_S dS_j\lbrack e e^{a\mu}
(4\Sigma^{bcj}T_{bc\mu}-\delta^j_\mu \Sigma^{bcd}T_{bcd})\rbrack\;,
\label{6}
\end{equation}
is $a$ component of the  gravitational energy-momentum flux, and

\begin{equation}
\Phi^a_m={1\over c}\int_S dS_j\,(ee^a\,_\mu T^{j\mu})\,,
\label{7}
\end{equation}
is the $a$ component of the matter energy-momentum flux. $S$
represents the spatial boundary of the volume $V$. Therefore the
loss of gravitational energy $E=cP^{(0)}$ is governed by the equation

\begin{equation}
{d\over {dt}}\biggl({E\over c^2}\biggr)=-\Phi^{(0)}_g-\Phi^{(0)}_m\;.
\label{8}
\end{equation}

The equations above have been applied to the evaluation of the
energy flux of plane and Einstein and Rosen waves \cite{Maluf4},
and to the analysis of loss of gravitational energy in Bondi and
Vaidya space-times \cite{Maluf5}. The results of these analyses
indicate that the
interpretation of Eqs. (6) and (7) as flux equations is correct.
In particular, Bondi's equation for the loss of mass is obtained
directly from Eq. (8) \cite{Maluf5}. It must be
emphasized that the continuity equation (5) is a direct
consequence of Einstein's equations (3).

In what follows we will construct (i) the tetrad field for the
Schwarzschild metric tensor adapted to a static observer,
(ii) the tetrad field adapted to a
moving observer in flat space-time, (iii) and the tetrad field for
a moving observer in the Schwarzschild space-time.

The Schwarzschild metric tensor in isotropic coordinates is given
by

\begin{equation}
ds^2=-A^2(dx^0)^2+B^2(d\rho^2 +\rho^2d\theta^2+
\rho^2\sin^2\theta d\phi^2)\;,
\label{9}
\end{equation}
where

\begin{eqnarray}
A^2&=& {{(1-m/2\rho)^2} \over {(1+m/2\rho)^2}}\;, \nonumber \\
B^2&=&(1+m/2\rho)^4 \;,  
\label{10}
\end{eqnarray}
and $m=MG/c^2$. The variable $\rho$ is related to the usual radial
coordinate $r$ by $r=\rho(1+m/2\rho)^2$. We define $x,y$ and
$z$ coordinates according to $x=\rho \sin\theta \cos\phi$,
$y=\rho \sin\theta \sin\phi$ and $z=\rho \cos\theta$. In terms of
these coordinates the metric tensor above is written as

\begin{equation}
ds^2=-A^2(dx^0)^2+B^2(dx^2+dy^2+dz^2)\;.
\label{11}
\end{equation}

The tetrad field $e^a\,_\mu$ maps coordinate differentials $dx^\mu$
of the physical space-time to coordinate differentials $dq^a$ that
locally describe a flat, reference space-time, according to
$dq^a=e^a\,_\mu dx^\mu$ \cite{Maluf1,Maluf4}. If the latter relation
can be globally integrated, the transformation is called holonomic
and both sets of coordinates, $q^a$ and $x^\mu$, describe globally
the flat space-time. In this case the tetrad field is simply given by
gradient vectors of the type $e^a\,_\mu=\partial q^a/\partial x^\mu$.
A nontrivial manifestation of the gravitational field
amounts to an anholonomic transformation between coordinate
differentials of the physical and reference space-times, in which
case $dq^a=e^a\,_\mu dx^\mu$ cannot be globally integrated and
$T^a\,_{\mu\nu} \ne 0$.

As discussed in Ref. \cite{Maluf1}, the tetrad fields for the
flat space-time that in
Cartesian coordinates satisfy the properties

\begin{eqnarray}
e_{(i)j}&=&e_{(j)i}\;, \\
e_{(i)}\,^0&=&0\;,
\label{12,13}
\end{eqnarray}
establish a unique reference space-time that is neither related by
a boost transformation nor rotating with respect to the physical
space-time. These conditions fix six degrees of freedom of 
$e^a\,_\mu$, and lead to $e^a\,_\mu=\delta^a_\mu$.
Although these conditions were derived for flat space-time
tetrad fields, we take them to hold for arbitrary
space-times. We can thus assert that the tetrad fields that satisfy
Eqs. (12) and (13) are adapted to static observers in space-time.
The tetrad fields for which Eqs. (12) and (13) hold and that yields
Eq. (11) read

\begin{equation}
e^a\,_\mu(x^0,x,y,z)=\pmatrix{A&0&0&0 \cr
0&B&0&0 \cr
0&0&B&0 \cr
0&0&0&B \cr}\;.
\label{14}
\end{equation}
($a$ and $\mu$ represent line and column indices, respectively).

Next, let us consider a boost transformation in flat space-time
determined by $q^{(0)}=\gamma(x^0-\beta x)$,
$q^{(1)}=\gamma(x-\beta x^0)$, $q^{(2)}=y$, $q^{(3)}=z$, where
$\gamma=\sqrt{1-\beta^2}$ and $\beta=v/c$. This transformation
yields the tetrad fields $e^a\,_\mu = \partial q^a/\partial x^\mu$
given by

\begin{equation}
e^a\,_\mu(x^0,x,y,z)=\pmatrix{\gamma& -\beta \gamma &0&0 \cr
-\beta \gamma & \gamma &0&0 \cr
0&0&1&0 \cr
0&0&0&1 \cr}\;.
\label{15}
\end{equation}
A similar set of tetrad fields has been recently employed 
by Mashhoon \cite{Mashhoon2} in the analysis of the electrodynamics
of linearly accelerated frames.
The right hand side of the equation above can be thought as the
transformation $e^a\,_\mu=\Lambda^a\,_b (e^b\,_\mu)_{flat}=
\Lambda^a\,_b (\delta^a_\mu)$.
Therefore in order to obtain the tetrad field for a moving observer
in the Schwarzschild space-time, that asymptotically undergoes a
Lorentz boost, we just multiply Eq. (14)
by the matrix $\Lambda^a\,_b$. We obtain

\begin{equation}
e^a\,_\mu(x^0,x,y,z)=\pmatrix{\gamma A& -\beta\gamma B&0&0 \cr
-\beta\gamma A&\gamma B&0&0 \cr
0&0&B&0 \cr
0&0&0&B \cr}\;.
\label{16}
\end{equation}
We remark that the expression above yields the metric
tensor Eq. (11) even for a time dependent frame velocity $v(t)$.
For a static object in
space-time whose four-velocity is given by $V^\mu=(1,0,0,0)$ we
may compute its frame components $V^a=e^a\,_\mu V^\mu=
(\gamma A,-\beta \gamma A,0,0)$.
Equation (16) is crucial for our analysis. Out of the latter we
calculate all components $T_{a\mu\nu}$  and 
$T_{\lambda \mu \nu}=e^a\,_\lambda T_{a\mu\nu}$.

The total gravitational energy $E$ as measured in the moving frame
follows from the $a=(0)$ component of $P^a$ calculated out of Eq.
(16),

\begin{equation}
P^{(0)}=-\int_{V\rightarrow \infty} d^3x\partial_j \Pi^{(0)j}=
4k\int_{S\rightarrow \infty} dS_j \,e e^{(0)}\,_0\Sigma^{00j}\;,
\label{17}
\end{equation}
where $e=AB^3$ and $e^{(0)}\,_0=\gamma A$. After a long but simple
calculation we obtain

\begin{equation}
\Sigma^{00j}=-{1\over {A^2B^3}}\partial_j B\;.
\label{18}
\end{equation}
$\Sigma^{00j}$ is obtained by taking all indices in Eq. (2)
to be space-time indices.
Evaluation of the integrand above in the limit
$\rho \rightarrow \infty$ (and identifying $\rho \simeq r$
in this limit) yields

\begin{equation}
dS_j\,ee^{(0)}\,_0 \Sigma^{00j}\simeq \gamma m  {1\over r^3}
(x\,dS_x+y\,dS_y+z\,dS_z)=\gamma m \sin\theta d\theta d\phi\;.
\label{19}
\end{equation}
After substituting the values of the constants $k$ and $m$ in Eqs.
(17) and (19) and integrating over a surface $S\rightarrow \infty$
we arrive at

\begin{equation}
P^{(0)}={E\over c}= \gamma M c\;.
\label{20}
\end{equation}

The only nonvanishing momentum component is given by

\begin{equation}
P^{(1)}=-\int_{V\rightarrow \infty} d^3x\partial_j \Pi^{(1)j}=
4k\int_{S\rightarrow \infty} dS_j \,e e^{(1)}\,_0\Sigma^{00j}\;,
\label{21}
\end{equation}
where $e^{(1)}\,_0=-\beta \gamma A$. Now we have

\begin{equation}
e\,e^{(1)}\,_0\Sigma^{00j}=\beta \gamma \partial_jB
\simeq -\beta \gamma m {x^j\over r^3}\;.
\label{22}
\end{equation}
It follows that

\begin{equation}
dS_j\,e e^{(1)}\,_0\Sigma^{00j}\simeq -\beta\gamma m\sin\theta
d\theta d\phi\;,
\label{23}
\end{equation}
in the limit $\rho \rightarrow \infty$.  After evaluating Eq. (21)
in this limit we find

\begin{equation}
P^{(1)}=-\beta \gamma M c=-{{Mv}\over \sqrt{1-v^2/c^2}}\;.
\label{24}
\end{equation}

The four-velocity of the moving frame is defined to be

$$u^a=(\gamma A,-\beta \gamma A, 0,0)\;,$$
which yields $u^au^b\eta_{ab}=-A^2$.
In terms of the four-velocity we may write

\begin{equation}
P^a=\biggl( {E\over c}, {\bf P}\biggr)=Mcu^a\;,
\label{25}
\end{equation}
if the moving frame is sufficiently far from the black hole
such that we can make  $A\simeq 1$.
We note that the square of $P^a$ yields

\begin{equation}
P^aP^b\eta_{ab}=-M^2c^2\;,
\label{26}
\end{equation}
irrespective of the value of $A$.
The similarity between black holes and elementary particles in
the expressions above is interesting. $P^a$ given by Eq. (25) is
the energy-momentum four-vector of a relativistic
particle of inertial mass $M$ \cite{Landau}.

The asymptotically flat limit of the tetrad field Eq. (16)
($A\rightarrow 1$ and $B\rightarrow 1$ in the limit
$\rho \rightarrow \infty$) is given by Eq. (15). Therefore in the
present analysis the usual asymptotically flat limit,

$$
e_{a\mu} \simeq \eta_{a\mu}+{1 \over2}h_{a\mu}({1\over r})\;,
$$
is not verified. This fact explains the emergence of the factor
$\gamma$ in Eq. (20). Thus for the moving frame in consideration
the total gravitational energy of the space-time differs by a
factor of $\gamma$ from the usual ADM value.
From the point of view of the TEGR, there is a certain
similarity between the concepts of mass in the special and
general theory of relativity. In particular, we observe that the
quantity $M$ that appears in Eq. (10) is the gravitational mass,
whereas in Eqs. (25) and (26), in the framework of special
relativity, $M$ is typically an inertial mass.

We mentioned earlier that $e^a\,_\mu$ given by Eq. (16) yields
the metric tensor Eq. (11) even for a time dependent velocity
$v(t)$. It turns out that if we assume a time dependent velocity
the calculations that lead to Eqs. (20), (24) and (25) remain
unchanged. In this case it may be verified after long
calculations that the total gravitational energy flux
$\Phi^{(0)}_g$ given by Eq. (6) is nonvanishing. By integrating over
a surface of constant $\rho$ and making $\rho \rightarrow \infty$
we find

\begin{eqnarray}
\Phi^{(0)}_g&=&
k\int_{S\rightarrow \infty} dS_j \biggl[ e e^{(0)}\,_0 g^{00}
(4\Sigma^{bcj}T_{bc0}-\delta^j_0 \Sigma^{bcd}T_{bcd}) \nonumber \\
&+& e e^{(0)}\,_1 g^{11}
(4\Sigma^{bcj}T_{bc1}-\delta^j_1 \Sigma^{bcd}T_{bcd})\biggr]
\nonumber \\
&\simeq& -k\beta\gamma \biggl[
\int_{S\rightarrow \infty} dS_x(2T_{001}T_{212}
+2T_{001}T_{313}) \nonumber \\
&+& \int_{S\rightarrow \infty} dS_y(2T_{001}T_{112}
-2T_{001}T_{323}) \nonumber \\
&+& \int_{S\rightarrow \infty} dS_z(2T_{001}T_{113}
+2T_{001}T_{223}) \biggr]\nonumber \\
&=&-k\beta\gamma \biggl[
12{{d \beta}\over {dx^0}}\gamma^2 m\int_0^{2\pi}d\phi\cos^2\phi
\int_0^\pi d\theta \sin^3\theta \biggr] \nonumber \\
&=& -{v\over c^2}\gamma^3\,M {{dv}\over {dt}}\;.
\label{27}
\end{eqnarray}
In the equation above the three surface integrals contribute
equally to the final result.
Alternatively, we can simply calculate the time derivative of
$P^{(0)}$ in order to obtain

\begin{equation}
{d \over {dx^0}}\lbrack \gamma(t) Mc\rbrack
={v\over c^2} \gamma^3\,M {{dv}\over {dt}}\;.
\label{28}
\end{equation}

Assuming a positive acceleration $dv/dt >0$, we see from Eq. (28)
that an observer in the accelerated frame would measure an
increasing value of the gravitational energy. This positive
variation of $E$ would be compensated by a negative flux of 
gravitational energy in this frame. Thus the gravitational energy
flux is a consequence of the gravitational energy conservation.

Therefore the moving observer is expected to experience in his
frame gravitational energy radiation, in similarity to the expected
phenomena in classical electrodynamics. Of course this is an
intricate conceptual issue in general relativity that deserves a
thorough investigation, but the situation
here is no different than in electrodynamics, except that in the
latter context one normally considers a static observer and an
accelerated charged particle. The gravitational
radiation emitted by accelerated black holes has been investigated
already by Tomimatsu \cite{Tomimatsu}, who considered black holes
with a more intricate structure. By allowing the observer to be
accelerated we may restrict considerations to a simple
Schwarzschild black hole. We remark that uniformly accelerated
particles described in the context of boost-rotation-symmetric
space-times \cite{Bicak} exhibit radiative character in the sense
of Bondi's approach to the radiation of isolated sources.
Likewise, the work
of Ref. \cite{Tomimatsu} consists of bringing the vacuum Weyl C
metric to Bondi's form of the metric tensor.

Finally we take the time derivative of the gravitational momentum
$P^{(1)}$ given by Eq. (24). It reads

\begin{equation}
{{dP^{(1)}\over {dt}}}=-{M \over {(1-v^2/c^2)^{3\over 2}}}
{{dv}\over {dt}}\;.
\label{29}
\end{equation}
A first interpretation for the equation above is that it represents
the force that would act on the black hole
(which is the source of the gravitational energy-momentum)
in order to describe its motion from the standpoint
of an observer at rest in the accelerated frame.
From this point of view, $M$ is indeed an inertial mass. However,
assuming again a positive acceleration, we see that a
negative variation of $P^{(1)}$ is compensated by a
positive flux of momentum as measured in the moving frame.

It must be noted that the definition of the gravitational
energy-momentum $P^{a}$, Eq. (4), is not equivalent to the
traditional ADM energy-momentum. Let us denote the latter by
$P^\mu$ (bearing in mind that it has no relationship to $P^{a}$).
The expression for the
gravitational ADM momenta $P^i$ is given by Eqs. (7.6.22) of Ref.
\cite{Weinberg} or Eq. (4.2) of Ref. \cite{Goldberg}. Since the
space-time described by Eq. (9) is static, it is clear that the
ADM momenta $P^i$ vanishes, in contrast to Eq. (24). Considering
$P^\mu$ to be a four-vector defined in the asymptotic region of the
space-time (it may be taken either as the asymptotic limit of a
pseudo-tensor \cite {Weinberg} or a surface - boundary - term of
the total Hamiltonian of an asymptotically flat space-time),
then by means of a transformation $\tilde x^\mu
=\Lambda^\mu\,_\nu x^\nu$ it is possible to obtain an expression
for $\tilde P^i$ similar to Eq. (25). For this purpose we must
interpret the latter transformation both as a general coordinate
and a SO(3,1) transformation. However, we note that
$P^a$ is a true SO(3,1) energy-momentum four-vector in the whole
space-time (the volume of integration $V$ in Eq. (4) may be
finite, although we have integrated over the whole spacelike
hypersurface). The frame transformation from Eq. (15) to Eq. (16)
is also defined in the whole space-time. $P^a$ is not restricted to
be a vector in the asymptotic region, and in general it is defined
for space-times with arbitrary boundary conditions. It can be
constructed for the de Sitter or anti-de Sitter spaces, for
instance.

In conclusion, we have seen that the gravitational energy-momentum
for a black hole of mass $M$ satisfies the usual relativistic
expressions for the energy-momentum of a relativistic particle
of mass $M$. Moreover, if the observer accelerates with respect to
the black hole, a gravitational energy flux is expected to
be nonvanishing in the accelerated frame. It follows from the
present analysis that the gravitational mass $M$ that appears in
Eqs. (9) and (10) may be taken to be an inertial mass in the
context of Eq. (29). At last, we observed that the gravitational
energy-momentum considered here may provide a unified picture of
the concept of mass-energy in the special and general theory of
relativity. If $M$ is the mass of a body in a rest frame, in a
moving frame its mass is increased by the usual factor $\gamma$,
as we we know from the special theory of relativity. The results
obtained here show that the increase of the value of the mass of
the body is due to gravitational radiation that takes place in
the passage from the rest frame to the moving frame, in which
case the motion must be accelerated.

\end{document}